\documentclass[aps,prd,10pt,showpacs]{revtex4}
\usepackage[latin1]{inputenc}
\usepackage{amsmath}
\usepackage{amsfonts}
\usepackage{amssymb}
\usepackage{amsthm}
\usepackage{bm}
\usepackage{mathrsfs}
\usepackage{tensor}
\usepackage{graphics}

\begin{document}

\title[Covariant Transformation Optics]{A Completely Covariant Approach to Transformation Optics}
\author{Robert T. Thompson}
\email{robert@cosmos.phy.tufts.edu}
\author{Steven A. Cummer$^{\dag}$}
\author{J\"org Frauendiener$^{\ddag}$}
\affiliation{$^{*\,\ddag}$ Department of Mathematics and Statistics,
University of Otago, P.O.\ Box 56, Dunedin, 9054,  New Zealand}
\affiliation{$^{\dag}$ Department of Electrical and Computer Engineering and Center for Metamaterials
and Integrated Plasmonics, Duke University, Durham, North Carolina 27708, USA}
\affiliation{$^{\ddag}$ Centre of Mathematics for Applications, University of Oslo, P.O.\ Box 1053, Blindern, NO-0316 Oslo, Norway}

\begin{abstract}
We show that the Plebanski based approach to transformation optics overlooks some subtleties in the electrodynamics of moving dielectrics that restricts its applicability to a certain class of transformations.  An alternative, completely covariant, approach is developed that is more generally applicable and provides a clearer picture of transformation optics.
\end{abstract}

\pacs{02.40.Hw, 41.20.Jb, 42.15.Eq, 42.25.-p}

\maketitle

\section{Introduction}
The concept of transformation optics enables coordinate transformations on electromagnetic fields to be physically realized as complex materials.  This idea, applied to electromagnetic fields, originated with Pendry \cite{Pendry:2006a}, who pointed out the specific relationship between spatial transformations and material properties, and demonstrated how it could be used to create novel devices.  Earlier, closely related work by Greenleaf et al. \cite{Greenleaf:2003} developed a similar concept for electric current flow and applied it to impedance tomography.  The original descriptions of transformation optics were formulated for purely spatial transformations \cite{Milton:2006,Schurig:2006}.  Based on a theoretical foundation developed by Plebanski \cite{Plebanski:1959ff} and De Felice \cite{DeFelice:1971}, Leonhardt and Philbin \cite{Leonhardt:2006a} described transformation optics in terms of differential geometry and provided a recipe for the material constitutive relations for transformations that involve both space and time.

However, the Plebanski/De Felice based formulation is non-covariant, and part of the interpretation ascribed by Leonhardt and Philbin overlooks some subtleties of electrodynamics in moving dielectrics, restricting its validity to a certain class of transformations.  These arise from a non-covariant matching of the constitutive relations in a curved, vacuum space-time with the constitutive relations for an \textit{isotropic} moving dielectric.  We show that such a matching cannot be made for a more general, anisotropic moving dielectric.  It is possible to make such an identification with a \textit{stationary} (i.e.\ at rest with respect to the frame in which the fields are defined) dielectric with non-vanishing magneto-electric coupling, but this can only be identified with a velocity if the material is isotropic.  Moreover, the applicability of these non-covariant formulations to complex initial materials is limited.  Although special cases have been addressed in the literature \cite{Cheng:2009}, a completely general formulation is lacking.

Here we present a completely covariant approach that is valid for all transformations, materials, and material motion.  Based on a modern, completely covariant, description of classical electrodynamics, it allows for a more rigorous and clear explanation of the concept behind, and interpretation of, transformation optics.  This approach attempts to provide a systematic framework for the study of transformation optics and could be especially useful for applications involving moving objects, general relativistic corrections.

The paper is organized as follows: In Sec.\ \ref{Sec:ClassElec} we review the completely covariant theory of electrodynamics in modern language.  Section \ref{Sec:MovingDielectrics} discusses electrodynamics in moving dielectrics and the differences between isotropic and anisotropic moving dielectrics; in particular how the non-covariant equations differ for isotropic and anisotropic materials.  Section \ref{Sec:Plebanski} outlines the Plebanski based approach, and explains the subtleties involved when taking into account the discussion of Sec.\ \ref{Sec:MovingDielectrics}.   Section \ref{Sec:TransformationOptics} describes the concept behind transformation optics and presents an interpretation consistent with the development of Sec.\ \ref{Sec:ClassElec}. The main results for applications in transformation optics are Eqs.\ (\ref{Eq:MaterialChi}) and (\ref{Eq:InitialVacMaterialChi}). A variety of examples are given in Sec.\ \ref{Sec:Examples} in order to both illustrate the details of the procedure, and to demonstrate agreement with known results obtained through other means.
We conclude with Sec.\ \ref{Sec:Conclusions} and discuss possible extensions of this work to more general materials.

\section{Modern Electrodynamics} \label{Sec:ClassElec}
In modern parlance, classical electrodynamics is formulated in very geometric terms.  In fact the homogeneous Maxwell equations are simply constraints on the fields imposed by the space-time geometry, and are a purely geometric condition.  The inhomogeneous vacuum equations arise quite naturally from an action principle and a dual operation possessed by a manifold equipped with a metric.  We follow the development and sign convention of Ref.\ \cite{Misner:1974qy}, and make use of the Einstein summation convention, where repeated indices are summed over.  The metric $g_{\alpha\beta}$ (or it's inverse $g^{\alpha\beta}$) lowers (raises) indices, and for the speed of light we use $c=1$.  For more details on the geometric foundations of electrodynamics the reader may consult a myriad of excellent sources such as \cite{Tu:2008,Bleecker,Baez}.

To make this more concrete, begin with the connection covector $\mathbf{A}=A_{\mu}$ (the covariant version of the contravariant 4-vector potential $A^{\mu}$), the \textit{field strength} $\mathbf{F}=F_{\mu\nu}$ is an antisymmetric tensor defined as the exterior derivative (a generalization of the differential of a function) of $\mathbf{A}$,
\begin{equation}
 \mathbf{F}=\mathrm{d}\mathbf{A}, \quad F_{\mu\nu} = A_{\nu,\mu}-A_{\mu,\nu}
\end{equation}
where the comma indicates a derivative. The components of $\mathbf{F}$ can be represented as a matrix, that for Minkowski space-time in Cartesian coordinates have values
\begin{equation} \label{Eq:FComponents}
 F_{\mu\nu} = \left(
 \begin{matrix}
  0 & -E_x & -E_y & -E_z\\
  E_x & 0 & B_z & -B_y\\
  E_y & -B_z & 0 & B_x\\
  E_z & B_y & -B_x & 0
 \end{matrix}
 \right).
\end{equation}

The field strength tensor $\mathbf{F}$ encodes some information about the fields, namely the electric field strength and the magnetic flux.  On the other side of this coin, the magnetic field strength and the electric flux are encoded in another tensor $\mathbf{G}$, called the \textit{excitation tensor}.  The components of $\mathbf{G}$ can also be represented as a matrix, that for Minkowski space-time in Cartesian coordinates have values
\begin{equation} \label{Eq:GComponents}
 G_{\mu\nu} = \left(
 \begin{matrix}
  0 & H_x & H_y & H_z\\
  -H_x & 0 & D_z & -D_y\\
  -H_y & -D_z & 0 & D_x\\
  -H_z & D_y & -D_x & 0
 \end{matrix}
 \right).
\end{equation}

In much of the existing literature the matrix expression given for $\mathbf{G}$ is not Eq.\ (\ref{Eq:GComponents}) but rather has $H_a$ and $D_a$ interchanged.  However, this obscures the physical nature of the fields, where, for example, the electric flux must be the purely spatial part of $\mathbf{G}$.  The origin of the difference in definitions of $\mathbf{G}$ and further implications of this difference will be discussed elsewhere \cite{Thompson:2010b}.  It is sufficient to know that, as the proper dual of $\mathbf{F}$, $\mathbf{G}$ exists on an equal tensor footing and transforms as a tensor rather than as a ``tensor density'', an unnecessary complication.  

The completely covariant form of Maxwell's equations are
\begin{equation} \label{Eq:HomogeneousMaxwell}
 \mathrm{d} \mathbf{F} = 0.
\end{equation}
and
\begin{equation} \label{Eq:InhomogeneousMaxwell}
 \mathrm{d}\mathbf{G}=\mathbf{J}
\end{equation}
where $\mathbf{J}=J_{\alpha\beta\gamma}$ is the covariant charge-current 3 index tensor.

There remains one piece to the electrodynamics puzzle: an equation describing how $\mathbf{F}$ and $\mathbf{G}$ are related.  The most natural way to define such a constitutive relation uses an operation $\star$, called the Hodge dual, that on a four dimensional manifold with a metric $\mathbf{g}=g_{\mu\nu}$ takes a covariant tensor and returns another covariant tensor such that
\begin{equation} \label{Eq:StarF}
 (\star \mathbf{F})_{\mu\nu} = \frac12 \sqrt{|g|}\,\epsilon_{\mu\nu\alpha\beta}g^{\alpha\gamma}g^{\beta\delta}F_{\gamma\delta},
\end{equation}
where $\epsilon_{\mu\nu\alpha\beta}$ is the totally antisymmetric Levi-Civita symbol. This means that $\star$ can be represented as a tensor with components
\begin{equation}
 \star\indices{_{\alpha\beta}^{\mu\nu}}=\frac12 \sqrt{|g|}\epsilon_{\alpha\beta\sigma\rho}g^{\sigma\mu}g^{\rho\nu},
\end{equation}
and contains information about the space-time.  The components of $\star \mathbf{F}$ can also be represented as a matrix, that for Minkowski space-time in Cartesian coordinates have values
\begin{equation} \label{Eq:StarFComponents}
 (\star \mathbf{F})_{\mu\nu} = \left(
 \begin{matrix}
  0 & B_x & B_y & B_z\\
  -B_x & 0 & E_z & -E_y\\
  -B_y & -E_z & 0 & E_x\\
  -B_z & E_y & -E_x & 0
 \end{matrix}
 \right).
\end{equation}
The constitutive equation is
\begin{equation} \label{Eq:Constitutive}
 \mathbf{G} = \boldsymbol{\chi}(\star\mathbf{F}),
\end{equation}
which in component form reads
\begin{equation} \label{Eq:ConstitutiveIndices}
 G_{\mu\nu} = \chi\indices{_{\mu\nu}^{\alpha\beta}}(\star\mathbf{F})_{\alpha\beta}.
\end{equation}

The tensor $\boldsymbol\chi$ contains information on the dielectric material's properties including permittivity, permeability, and magneto-electric couplings, and can be thought of as representing an averaging over all the material contributions to an action that describes a more fundamental quantum field theory.  To retain the desired properties and usual notions of $\mathbf{G}$ and $\mathbf{F}$, $\boldsymbol{\chi}$ must be independently antisymmetric on its first two and last two indices, and in vacuum $\boldsymbol{\chi}(\star\mathbf{F})=\star\mathbf{F}$.
Additional symmetry conditions are frequently imposed \cite{Landau:1960,Post:1962} for various reasons not considered here because we would like to be able to describe very general materials \cite{Yuan:2009,Popa:2007}, but the last condition is sufficient to uniquely specify all components of $\boldsymbol{\chi}$ for the vacuum, they are
\begin{equation} \label{Eq:VacChi}
 \boldsymbol{\chi}_{vac}=(\chi_{vac})_{\gamma\delta}^{\phantom{\gamma\delta}\sigma\rho} = \frac12\left(
\begin{matrix}
 \left(\begin{smallmatrix}
 0 & 0 & 0 & 0\\ 0 & 0 & 0 & 0\\ 0 & 0 & 0 & 0\\ 0 & 0 & 0 & 0\\
 \end{smallmatrix} \right) &
 \left(\begin{smallmatrix}
 0 & 1 & 0 & 0\\ -1 & 0 & 0 & 0\\ 0 & 0 & 0 & 0\\ 0 & 0 & 0 & 0\\
 \end{smallmatrix} \right) &
 \left(\begin{smallmatrix}
 0 & 0 & 1 & 0\\ 0 & 0 & 0 & 0\\ -1 & 0 & 0 & 0\\ 0 & 0 & 0 & 0\\
 \end{smallmatrix} \right) &
 \left(\begin{smallmatrix}
 0 & 0 & 0 & 1\\ 0 & 0 & 0 & 0\\ 0 & 0 & 0 & 0\\ -1 & 0 & 0 & 0\\
 \end{smallmatrix} \right) \\
 \left(\begin{smallmatrix}
 0 & -1 & 0 & 0\\ 1 & 0 & 0 & 0\\ 0 & 0 & 0 & 0\\ 0 & 0 & 0 & 0\\
 \end{smallmatrix} \right) &
 \left(\begin{smallmatrix}
 0 & 0 & 0 & 0\\ 0 & 0 & 0 & 0\\ 0 & 0 & 0 & 0\\ 0 & 0 & 0 & 0\\
 \end{smallmatrix} \right) &
 \left(\begin{smallmatrix}
 0 & 0 & 0 & 0\\ 0 & 0 & 1 & 0\\ 0 & -1 & 0 & 0\\ 0 & 0 & 0 & 0\\
 \end{smallmatrix} \right) &
 \left(\begin{smallmatrix}
 0 & 0 & 0 & 0\\ 0 & 0 & 0 & 1\\ 0 & 0 & 0 & 0\\ 0 & -1 & 0 & 0\\
 \end{smallmatrix} \right) \\
 \left(\begin{smallmatrix}
 0 & 0 & -1 & 0\\ 0 & 0 & 0 & 0\\ 1 & 0 & 0 & 0\\ 0 & 0 & 0 & 0\\
 \end{smallmatrix} \right) &
 \left(\begin{smallmatrix}
 0 & 0 & 0 & 0\\ 0 & 0 & -1 & 0\\ 0 & 1 & 0 & 0\\ 0 & 0 & 0 & 0\\
 \end{smallmatrix} \right) &
 \left(\begin{smallmatrix}
 0 & 0 & 0 & 0\\ 0 & 0 & 0 & 0\\ 0 & 0 & 0 & 0\\ 0 & 0 & 0 & 0\\
 \end{smallmatrix} \right) &
 \left(\begin{smallmatrix}
 0 & 0 & 0 & 0\\ 0 & 0 & 0 & 0\\ 0 & 0 & 0 & 1\\ 0 & 0 & -1 & 0\\
 \end{smallmatrix} \right) \\
 \left(\begin{smallmatrix}
 0 & 0 & 0 & -1\\ 0 & 0 & 0 & 0\\ 0 & 0 & 0 & 0\\ 1 & 0 & 0 & 0\\
 \end{smallmatrix} \right) &
 \left(\begin{smallmatrix}
 0 & 0 & 0 & 0\\ 0 & 0 & 0 & -1\\ 0 & 0 & 0 & 0\\ 0 & 1 & 0 & 0\\
 \end{smallmatrix} \right) &
 \left(\begin{smallmatrix}
 0 & 0 & 0 & 0\\ 0 & 0 & 0 & 0\\ 0 & 0 & 0 & -1\\ 0 & 0 & 1 & 0\\
 \end{smallmatrix} \right) &
 \left(\begin{smallmatrix}
 0 & 0 & 0 & 0\\ 0 & 0 & 0 & 0\\ 0 & 0 & 0 & 0\\ 0 & 0 & 0 & 0\\
 \end{smallmatrix} \right) \\
\end{matrix} \right).
\end{equation}
Equation (\ref{Eq:VacChi}) expresses $\boldsymbol{\chi}$ as a matrix of matrices, the first two indices of $\chi\indices{_{\alpha\beta}^{\mu\nu}}$ give the $\alpha\beta$ component of the large matrix, which is itself a matrix described by the second set of indices.  The component values of $\boldsymbol{\chi}$ \textit{for the vacuum} are unique and independent of coordinate system. For a more general material, the component values can easily be determined by simply matching the results of the constitutive equation $\mathbf{G}=\boldsymbol{\chi}(\star\mathbf{F})$ with the usual flat-space constitutive relations in a particular coordinate system.  The components of the constitutive equation provide a set of six independent equations that can locally be collected in the form
\begin{equation} \label{Eq:ConstitutiveComponents1}
 \vec{H}=\check{\mu}^{-1}\vec{B}+\check{\gamma_1}^*\vec{E}, \quad \vec{D}=\check{\varepsilon}^*\vec{E}+\check{\gamma_2}^*\vec{B}.
\end{equation}
where we use the notation $\check{a}$ to denote a $3\times 3$ matrix.  Rearranging these to 
\begin{equation} \label{Eq:ConstitutiveComponents2}
 \vec{B}=\check{\mu}\vec{H}+\check{\gamma_1}\vec{E}, \quad \vec{D}=\check{\varepsilon}\vec{E}+\check{\gamma_2}\vec{H}
\end{equation}
gives the more familiar representation for the constitutive relations.  These three-dimensional representations of the completely covariant Eq.\ (\ref{Eq:Constitutive}) are essentially equivalent, and it is a simple matter to switch between them using the relations
\begin{equation} \label{Eq:ConstitutiveShift}
 \check{\mu}=\left(\check{\mu}^{-1}\right)^{-1}, \quad \check{\varepsilon}=\check{\varepsilon}^*-\check{\gamma_2}^*\check{\mu}\check{\gamma_1}^*, \quad \check{\gamma_1}=-\check{\mu}\check{\gamma_1}^*, \quad \check{\gamma_2} = \check{\gamma_2}^*\check{\mu}.
\end{equation}
However, one should be aware that these $3\times 3$ matrices are not tensors but simply components of $\boldsymbol{\chi}$ that have been collected into matrices.  To recover the usual component relations of Eq.\ (\ref{Eq:ConstitutiveComponents1}) the constitutive equation $\mathbf{G}=\boldsymbol{\chi}(\star\mathbf{F})$ allows us to make the identification
\begin{equation} \label{Eq:CartesianChi}
 \chi\indices{_{\gamma\delta}^{\sigma\rho}} = \frac12\left(
\begin{matrix}
 \left(\begin{smallmatrix}
 0 & 0 & 0 & 0\\ 0 & 0 & 0 & 0\\ 0 & 0 & 0 & 0\\ 0 & 0 & 0 & 0\\
 \end{smallmatrix} \right) &

* &

* &

* \\[15pt]

 \left(\begin{smallmatrix}
 0 & -\mu^{-1}_{xx} & -\mu^{-1}_{xy} & -\mu^{-1}_{xz}\\ 
 \mu^{-1}_{xx} & 0 & -\gamma_{1xz} & \gamma_{1xy}\\ 
 \mu^{-1}_{xy} & \gamma_{1xz} & 0 & -\gamma_{1xx}\\ 
 \mu^{-1}_{xz} & -\gamma_{1xy} & \gamma_{1xx} & 0
 \end{smallmatrix} \right) &

 \left(\begin{smallmatrix}
 0 & 0 & 0 & 0\\ 0 & 0 & 0 & 0\\ 0 & 0 & 0 & 0\\ 0 & 0 & 0 & 0
 \end{smallmatrix} \right) &

* &

* \\[20pt]

 \left(\begin{smallmatrix}
 0 & -\mu^{-1}_{yx} & -\mu^{-1}_{yy} & -\mu^{-1}_{yz}\\ 
 \mu^{-1}_{yx} & 0 & -\gamma_{1yz} & \gamma_{1yy}\\ 
 \mu^{-1}_{yy} & \gamma_{1yz} & 0 & -\gamma_{1yx}\\ 
 \mu^{-1}_{yz} & -\gamma_{1yy} & \gamma_{1yx} & 0
 \end{smallmatrix} \right) &

 \left(\begin{smallmatrix}
 0 & -\gamma_{2zx} & -\gamma_{2zy} & -\gamma_{2zz}\\
 \gamma_{2zx} & 0 & -\epsilon_{zz} & \epsilon_{zy}\\
 \gamma_{2zy} & \epsilon_{zz} & 0 & -\epsilon_{zx}\\
 \gamma_{2zz} & -\epsilon_{zy} & \epsilon_{zx} & 0
 \end{smallmatrix} \right) &

 \left(\begin{smallmatrix}
 0 & 0 & 0 & 0\\ 0 & 0 & 0 & 0\\ 0 & 0 & 0 & 0\\ 0 & 0 & 0 & 0\\
 \end{smallmatrix} \right) &

* \\[20pt]

 \left(\begin{smallmatrix}
 0 & -\mu^{-1}_{zx} & -\mu^{-1}_{zy} & -\mu^{-1}_{zz}\\ 
 \mu^{-1}_{zx} & 0 & -\gamma_{1zz} & \gamma_{1zy}\\ 
 \mu^{-1}_{zy} & \gamma_{1zz} & 0 & -\gamma_{1zx}\\ 
 \mu^{-1}_{zz} & -\gamma_{1zy} & \gamma_{1zx} & 0
 \end{smallmatrix} \right) &

 \left(\begin{smallmatrix}
 0 & \gamma_{2yx} & \gamma_{2yy} & \gamma_{2yz}\\
 -\gamma_{2yx} & 0 & \epsilon_{yz} & -\epsilon_{yy}\\
 -\gamma_{2yy} & -\epsilon_{yz} & 0 & \epsilon_{yx}\\
 -\gamma_{2yz} & \epsilon_{yy} & -\epsilon_{yx} & 0
 \end{smallmatrix} \right) &

 \left(\begin{smallmatrix}
 0 & -\gamma_{2xx} & -\gamma_{2xy} & -\gamma_{2xz}\\
 \gamma_{2xx} & 0 & -\epsilon_{xz} & \epsilon_{xy}\\
 \gamma_{2xy} & \epsilon_{xz} & 0 & -\epsilon_{xx}\\
 \gamma_{2xz} & -\epsilon_{xy} & \epsilon_{xx} & 0
 \end{smallmatrix} \right) &

 \left(\begin{smallmatrix}
 0 & 0 & 0 & 0\\ 0 & 0 & 0 & 0\\ 0 & 0 & 0 & 0\\ 0 & 0 & 0 & 0
 \end{smallmatrix} \right) \\
\end{matrix} \right),
\end{equation}
where the $*$ indicates entries that are antisymmetric on either the first or second set of indices on $\chi\indices{_{\gamma\delta}^{\sigma\rho}}$.

\section{Electrodynamics of Moving Dielectrics} \label{Sec:MovingDielectrics}
Consider a dielectric material moving with uniform velocity with respect to the lab frame.  The material properties of the dielectric are known in the material frame -- the frame at rest with the material -- and are described by permittivity $\check{\varepsilon}$, permeability $\check{\mu}$ and magneto-electric couplings $\check{\gamma_1}$ and $\check{\gamma_2}$.  For a field tensor $\mathbf{F}$ measured in the lab frame, the components of $\mathbf{F}$ relative to the material frame are found by changing to the material frame with a Lorentz transformation, $\mathbf{L}$ \cite{Misner:1974qy,Jackson} 
\begin{equation}
 \mathbf{F}' = \mathbf{L}(\mathbf{F}) \quad \Rightarrow \quad F_{\alpha'\beta'}=L\indices{^{\alpha}_{\alpha'}}L\indices{^{\beta}_{\beta'}}F_{\alpha\beta}.
\end{equation}
Inside the material, $\mathbf{G}'=\mathbf{L}(\mathbf{G})$ is now related to $\mathbf{F}'$ by the usual constitutive equation $\mathbf{G}'=\boldsymbol{\chi}(\star\mathbf{F}')$, or
\begin{equation} \label{Eq:MovingRelations1}
 \mathbf{L}(\mathbf{G}) = \boldsymbol{\chi}\left(\mathbf{L}(\star\mathbf{F})\right),
\end{equation}
resulting in expressions involving combinations of $\vec{D}$ and $\vec{H}$ in terms of combinations of $\vec{E}$ and $\vec{B}$. On the other hand, the inverse transformation of $\mathbf{G}'$,
\begin{equation} \label{Eq:MovingRelations2}
 \mathbf{G} = \mathbf{L}^{-1}\left[\boldsymbol{\chi}\left(\mathbf{L}(\star\mathbf{F})\right)\right],
\end{equation}
gives direct expressions for the components forming $\vec{D}$ and $\vec{H}$. 
Furthermore, since the constitutive map $\mathbf{G}=\boldsymbol{\chi}(\star\mathbf{F})$ is totally covariant, a third representation can be obtained by transforming $\boldsymbol{\chi}$ from the material frame back to the lab frame
\begin{equation} \label{Eq:MovingRelations3}
 \mathbf{G} = \left(\mathbf{L}^{-1}(\boldsymbol{\chi})\right)(\star\mathbf{F}).
\end{equation}

Equations (\ref{Eq:MovingRelations1}) - (\ref{Eq:MovingRelations3}) provide equivalent representations for the constitutive relations in a moving dielectric, are the most convenient notation for dealing with electrodynamics in a 4-dimensional, covariant way, and are valid for arbitrary space-times and transformations.  If one insists on expressing these four dimensional relations in terms of 3-vectors, then for the special case of a dielectric material moving through the Minkowski vacuum with uniform velocity,
\begin{subequations} \label{Eq:MinkowskiMoving}
\begin{equation}
 \vec{H} + \left(\vec{D}\times\vec{\beta}\right) =
 \check{\mu}^{-1}\left[\vec{B} + \left(\vec{E}\times\vec{\beta}\right) \right] +
 \check{\gamma_1}^*\left[\vec{E}- \left(\vec{B}\times\vec{\beta}\right) \right],
\end{equation}
\begin{equation}
 \vec{D} - \left(\vec{H}\times\vec{\beta}\right) = 
 \check{\varepsilon}^*\left[\vec{E} - \left(\vec{B}\times\vec{\beta}\right) \right] + 
 \check{\gamma_2}^*\left[\vec{B} + \left(\vec{E}\times\vec{\beta}\right) \right].
\end{equation}
\end{subequations}

These equations are typically shown in the literature only for the restricted case where the material is isotropic, $\check{\varepsilon}=\varepsilon$, $\check{\mu}=\mu$, and has no magneto-electric coupling, $\check{\gamma_1}=\check{\gamma_2}=0$ (see, e.g.\ Ref.\ \cite{Landau:1960}).  For this restricted case, the low velocity limit is
\begin{subequations} \label{Eq:LowVelocityRelations}
 \begin{equation}
  \vec{D} = \varepsilon\vec{E}+(\varepsilon\mu-1)(\vec{\beta}\times \vec{H}),
 \end{equation}
 \begin{equation}
  \vec{B} = \mu\vec{H} - (\varepsilon\mu-1)(\vec{\beta}\times \vec{E}).
 \end{equation}
\end{subequations}

While these isotropic, low-velocity results may be the most familiar, they are not correct when $\check{\varepsilon}$ and $\check{\mu}^{-1}$ are matrix valued.  Indeed, following the same procedure leading from Eqs.\ (\ref{Eq:MinkowskiMoving}) to Eq.\ (\ref{Eq:LowVelocityRelations}) with matrix-valued $\check{\varepsilon}$ and $\check{\mu}^{-1}$ and with $\check{\gamma_1}=\check{\gamma_2}=0$, leads instead to
\begin{subequations} \label{Eq:CorrectLowVelocity}
 \begin{equation}
  \vec{D} = \check{\varepsilon}\vec{E} + \check{\varepsilon} \left(\vec{\beta}\times\check{\mu}\vec{H}\right) - \left(\vec{\beta}\times\vec{H}\right),
 \end{equation}
\begin{equation}
 \vec{B} = \check{\mu}\vec{H} + \check{\mu} \left(\check{\varepsilon}\vec{E}\times\vec{\beta}\right) - \left(\vec{E}\times\vec{\beta}\right).
\end{equation}
\end{subequations}
Matrix multiplication is not commutative with the cross product.  The difference in going from an isotropic material to a anisotropic material may seem trivial at first -- and this is true when one maintains covariance, as in Eq.\ (\ref{Eq:MinkowskiMoving}) -- but by giving up covariance in passing to the low-velocity limit the results change significantly.  Even diagonality of $\check{\varepsilon}$ and $\check{\mu}^{-1}$ is not sufficient, one may readily show that recovering Eqs.\ (\ref{Eq:LowVelocityRelations}) requires pure isotropy.

\section{Plebanski Based Transformation Optics} \label{Sec:Plebanski}
Plebanski \cite{Plebanski:1959ff} previously studied the propagation of electromagnetic waves in gravitational fields, and much of the current analysis in transformation optics is based on his results . Consider the propagation of electromagnetic waves in a vacuum space-time described by an arbitrary metric $g_{\alpha\beta}$. Using $g_{\alpha\beta}$ and $\boldsymbol{\chi}=\boldsymbol{\chi}_{vac}$ in Eq.\ (\ref{Eq:Constitutive}), the constitutive relations for this vacuum space-time can be written
\begin{subequations} \label{Eq:Plebanski1}
 \begin{equation}
  D_{a}=-\frac{\sqrt{|g|}}{g_{00}}g^{ab}E_b + \epsilon_{abc}\frac{g_{0b}}{g_{00}}H_c,
 \end{equation}
 \begin{equation}
  B_a = -\frac{\sqrt{|g|}}{g_{00}}g^{ab}H_b - \epsilon_{abc}\frac{g_{0b}}{g_{00}}E_c.
 \end{equation}
\end{subequations}
Written this way these equations are not, as cautioned by Plebanski, covariant.  Notice that while they still sum over repeated indices they do not conserve index type. Comparing with the results for a moving dielectric in the low velocity limit, Eqs.\ (\ref{Eq:LowVelocityRelations}), it would appear that a dielectric moving with low velocity through vacuum Minkowski space-time with permeability and permittivity
\begin{equation} \label{Eq:Plebanski2}
 \check{\varepsilon} = \check{\mu} = -\frac{\sqrt{|g|}}{g_{00}}g^{ab},
\end{equation}
and (scaled) velocity
\begin{equation}
 v_{b} = \frac{g_{0b}}{g_{00}},
\end{equation}
will reproduce the same field relations as the arbitrary vacuum space-time.

However, as we have seen in Sec.\ \ref{Sec:MovingDielectrics}, Eqs.\ (\ref{Eq:LowVelocityRelations}) are only valid when the material is isotropic, therefore this approach is not valid in general.  The problem stems from attempting to match the covariant Eqs.\ (\ref{Eq:Plebanski1}) with the non-covariant Eqs.\ (\ref{Eq:LowVelocityRelations}).  It is clear that such a matching cannot be made with the more general Eqs.\ (\ref{Eq:CorrectLowVelocity}).  This method, therefore, only works when the material is isotropic (in which case Eqs.\ (\ref{Eq:CorrectLowVelocity}) reduces to Eq.\ (\ref{Eq:LowVelocityRelations})), or when $g_{0b}=0$ for all $b$ (in which case the velocity must be assumed zero).

Despite these restrictions, this approach frequently gives, or appears to give, correct results. This is because there \textit{is} a case where the covariant Eqs.\ (\ref{Eq:Plebanski1}) can be matched with a covariant expression in dielectrics.  Remember that the general constitutive relations for a \textit{stationary} (at rest with respect to the frame in which the fields are defined) dielectric can be written in the form of either Eqs.\ (\ref{Eq:ConstitutiveComponents1}) or Eqs.\ (\ref{Eq:ConstitutiveComponents2}), which are related by Eqs.\ (\ref{Eq:ConstitutiveShift}).  Matching Eqs.\ (\ref{Eq:ConstitutiveComponents2}) with Eqs.\ (\ref{Eq:Plebanski1}), one could identify $\check{\varepsilon}$ and $\check{\mu}$ as in Eq.\ (\ref{Eq:Plebanski2}) but with magneto-electric coupling
\begin{equation}
 \check{\gamma_2}^{\mathtt{T}} = \check{\gamma_1}= -\epsilon_{abc}\frac{g_{0b}}{g_{00}}.
\end{equation}

It is important to realize that when trying to match Eqs.\ (\ref{Eq:Plebanski1}) with Eq.\ (\ref{Eq:ConstitutiveComponents2}), that the $\vec{E}$, $\vec{B}$, etc.\ found in Eq.\ (\ref{Eq:ConstitutiveComponents2}) are not actually vectors but components of either 1-forms or 2-forms, and that interchanging a 1-form with a vector is only trivial in Minkowski space-time.  It is also important to realize that this magneto-electric coupling cannot always be simply interpreted as a velocity.  Indeed, returning to the more general Eqs.\ (\ref{Eq:MinkowskiMoving}) one can find the low velocity limits (again in Minkowski space-time)
\begin{subequations}
\begin{equation}
 \vec{D} = \check{\varepsilon}\vec{E} + \check{\gamma_2}\vec{H} - \check{\varepsilon}\left(\check{\mu}\vec{H}\times\vec{\beta}\right) + \left(\vec{H}\times\vec{\beta}\right) - \check{\varepsilon}\left(\check{\gamma_1}\vec{E}\times\vec{\beta}\right) + \check{\gamma_2}\left(\check{\varepsilon}\vec{E}\times\vec{\beta}\right) + \check{\gamma_2}\left(\check{\gamma_2}\vec{H}\times\vec{\beta}\right),
\end{equation}
\begin{equation}
 \vec{B} = \check{\mu}\vec{H} + \check{\gamma_1}\vec{E} + \check{\mu}\left(\check{\varepsilon}\vec{E}\times\vec{\beta}\right) - \left(\vec{E}\times\vec{\beta}\right) + \check{\mu}\left(\check{\gamma_2}\vec{H}\times\vec{\beta}\right)- \check{\gamma_1}\left(\check{\mu}\vec{H}\times\vec{\beta}\right)  -\check{\gamma_1}\left(\check{\gamma_1}\vec{E}\times\vec{\beta}\right),
\end{equation}
\end{subequations}
from which it should be clear that the only way to identify a magneto-electric coupling with a velocity is to replace a stationary magneto-electric material with an \textit{isotropic} moving dielectric with no magneto-electric coupling.  Instead, it is better to employ a fully covariant procedure based on the theory as developed thus far, which is not only valid for a very general class of materials and arbitrary dielectric motions \cite{Cheng:2009}, but more clearly distinguishes between the material and space-time contributions, and is also valid for arbitrary background space-times, such as the weakly curved space-time around Earth.

\section{Transformation Optics} \label{Sec:TransformationOptics}
Due to the apparent limitations and non-covariance of the Plebanski/De Felice based approach to transformation optics, we ask whether there is a more general approach.  By following the covariant development of electrodynamics in the previous sections we find that there is.  In an attempt to more clearly tie the concept to the method, consider the following interpretation of transformation optics.

Start with an initial space-time manifold $(M,\mathbf{g},\star)$, field configuration $(\mathbf{F},\mathbf{G},\mathbf{J})$, and material distribution $\boldsymbol{\chi}$, where $\mathrm{d}\mathbf{F}=0$, $\mathrm{d}\mathbf{G}=\mathbf{J}$, and $\mathbf{G}=\boldsymbol{\chi}(\star\mathbf{F})$.  Imagine now a map $T:M\to \tilde{M}\subseteq M$ that maps $M$ to some image $\tilde{M}$ and transforms the electromagnetic fields in some smooth way to a new configuration $(\tilde{\mathbf{F}},\tilde{\mathbf{G}},\tilde{\mathbf{J}})$.  Because the underlying space-time is physically unaltered the manifold is still described by $(M,\mathbf{g},\star)$.  But for the new field configuration to be physically supported there must exist a new material distribution $\tilde{\boldsymbol{\chi}}$.  Thus on the image $\tilde{M}$ we must have $\mathrm{d}\tilde{\mathbf{F}}=0$, $\mathrm{d}\tilde{\mathbf{G}}=\tilde{\mathbf{J}}$, and $\tilde{\mathbf{G}}= \tilde{\boldsymbol{\chi}}(\star\tilde{\mathbf{F}})$.  Such a transformation could, for example, map $M$ to an image $\tilde{M}$ that contains a hole, i.e.\ a region from which the fields will be excluded, as in the case of an electromagnetic cloak \cite{Pendry:2006a,Rahm:200887}.   

Using the inverse $\mathcal{T}$ of the map $T$ (but see the remarks below), we can relate the initial $\mathbf{F}$ and $\mathbf{G}$ to the final $\tilde{\mathbf{F}}$ and $\tilde{\mathbf{G}}$ by an operation called the \textit{pullback} of $\mathcal{T}$, which we denote as $\mathcal{T}^*$.  Schematically, this means that we have the relation
\begin{equation}
 \tilde{\mathbf{G}} = \mathcal{T}^*(\mathbf{G}) = \mathcal{T}^*(\boldsymbol{\chi}(\star\mathbf{F})) = \tilde{\boldsymbol{\chi}}(\star\mathcal{T}^*(\mathbf{F})).
\end{equation}
This can be solved for $\tilde{\boldsymbol{\chi}}$ as a function of $x\in\tilde{M}$ to find \cite{Thompson:2010b}
\begin{equation} \label{Eq:MaterialChi}
 \tilde{\chi}\indices{_{\lambda\kappa}^{\tau\eta}}(x)=-\Lambda\indices{^{\alpha}_{\lambda}} \Lambda\indices{^{\beta}_{\kappa}} \chi\indices{_{\alpha\beta}^{\mu\nu}}\Big|_{\mathcal{T}(x)} \star\indices{_{\mu\nu}^{\sigma\rho}}\Big|_{\mathcal{T}(x)} (\Lambda^{-1})\indices{^{\pi}_{\sigma}}(\Lambda^{-1})\indices{^{\theta}_{\rho}}\, \star\indices{_{\pi\theta}^{\tau\eta}}
\end{equation}
In Eq.\ (\ref{Eq:MaterialChi}) $\boldsymbol{\Lambda}$ is the Jacobian matrix of $\mathcal{T}$, $\boldsymbol{\Lambda}^{-1}$ is the matrix inverse of $\boldsymbol{\Lambda}$, both $\boldsymbol{\Lambda}$ and $\boldsymbol{\Lambda}^{-1}$ are evaluated at $x$, and in solving for $\tilde{\boldsymbol{\chi}}$ we have made use of the fact that on a 4-dimensional Lorentzian manifold, acting twice with $\star$ returns the negative, $\star\star\mathbf{F}=-\mathbf{F}$.

Note that the initial material parameters must be evaluated at $\mathcal{T}(x)$, but if the initial space-time is vacuum, then since $\boldsymbol{\chi}_{vac}\star = \star$,
\begin{equation} \label{Eq:InitialVacMaterialChi}
 \tilde{\chi}\indices{_{\lambda\kappa}^{\tau\eta}}(x)=-\Lambda\indices{^{\alpha}_{\lambda}} \Lambda\indices{^{\beta}_{\kappa}}  \star\indices{_{\alpha\beta}^{\sigma\rho}}\Big|_{\mathcal{T}(x)} (\Lambda^{-1})\indices{^{\pi}_{\sigma}}(\Lambda^{-1})\indices{^{\theta}_{\rho}}\, \star\indices{_{\pi\theta}^{\tau\eta}},
\end{equation}
where one $\star$ is evaluated at $\mathcal{T}(x)$, and everything else is evaluated at $x$.  The prescriptions of Eqs.\ (\ref{Eq:MaterialChi}) or (\ref{Eq:InitialVacMaterialChi}) are meaningful only for points $x\in\tilde{M}$.  So for transformations such as the electromagnetic cloak, where there is a hole in $\tilde{M}$, the material parameters inside the hole are unspecified and completely arbitrary.  In this way, any uncharged material may be hidden inside the cloak without affecting the behavior of the fields outside.

Equation (\ref{Eq:MaterialChi}) is the core of transformation optics.  Start with a given space-time with metric $\mathbf{g}$ and associated dual $\star$, and with given dielectric material properties described by the tensor $\boldsymbol{\chi}$.  The initial space-time may be Minkowski and the initial dielectric may be the vacuum, but this is not necessary.  We imagine a transformation that changes the fields in some way. We ask what $\tilde{\boldsymbol{\chi}}$ is required to physically achieve such a transformation.  The answer is given by Eq.\ (\ref{Eq:MaterialChi}).

Two final remarks are in order.  One is to note that the crucial map is not $T$, but rather $\mathcal{T}$.  So one could just as well start with some map that here we label $\mathcal{T}$, and completely ignore $T$.  This is important because a given $T$ may not have an inverse, so we must assume that the given $\mathcal{T}$ is well defined on its domain, $\tilde{M}$.  But for practical purposes $\mathcal{T}$ can be thought of as the inverse of $T$.  Secondly, while Eqs.\ (\ref{Eq:MaterialChi}) and (\ref{Eq:InitialVacMaterialChi}) would be quite difficult to evaluate by hand, a modern computer algebra package makes the calculations almost trivial.

\section{Examples} \label{Sec:Examples}
The completely covariant approach to transformation optics developed above provides a concrete and powerful framework for analysing any desired configuration of fields and linear dielectric materials in any space-time and with any relative velocities.  This section presents some examples to illustrate the usefulness of the covariant approach described above and show that it recovers previous results obtained through other means.  In these examples we use a notation whereby the point $(t',x',y',z')$ is a point in the original, vacuum, space-time, while a point $(t,x,y,z)$ is a point in the space-time plus material.  We seek to describe the material properties as a function of the unprimed coordinates, $\boldsymbol{\chi}(t,x,y,z)$.

\subsection{Spatial Inversion}
For a simple inversion of one spatial coordinate, the transformation
\begin{equation}
 T(t',x',y',z')=(t,x,y,z)=(t',-ax',y',z')
\end{equation}
forms the basis for a ``superlens'' \cite{PhysRevLett.85.3966,Leonhardt:2006ai} and has the inverse map
\begin{equation}
 \mathcal{T}(t,x,y,z)=(t',x',y',z')=\left(t,-\frac{x}{a},y,z\right).
\end{equation}
The Jacobian matrix of $\mathcal{T}$ is $\Lambda\indices{^{\alpha}_{\beta}}=diag(1,-1/a,1,1)$.  Using this in Eq.\ (\ref{Eq:InitialVacMaterialChi}) results in a $\tilde{\boldsymbol{\chi}}$, that, upon comparison with Eq.\ (\ref{Eq:CartesianChi}), leads to
\begin{equation}
 \check{\varepsilon} = \check{\mu} = 
 \begin{pmatrix}
  -a & 0 & 0\\
  0 & -\frac1a & 0\\
  0 & 0 & -\frac1a
 \end{pmatrix}, \quad \check{\gamma_1}=\check{\gamma_2}=0.
\end{equation}

\subsection{Temporal Inversion}
On the other hand, inverting the time coordinate with the transformation
\begin{equation}
 T(t',x',y',z')=(t,x,y,z)=(-at',x',y',z')
\end{equation}
leads to
\begin{equation} \label{Eq:TimeInversion}
 \check{\varepsilon} = \check{\mu} = 
 \begin{pmatrix}
  -a & 0 & 0\\
  0 & -a & 0\\
  0 & 0 & -a
 \end{pmatrix}.
\end{equation}
Of course, replacing $a\to -a$ in the previous example corresponds to a simple stretching of the time coordinate, and the result is the same as Eq.\ (\ref{Eq:TimeInversion}) with $a\to -a$.  Both this example and the previous highlight the ease with which the completely covariant method handles coordinate inversions -- a concern raised in Ref.\ \cite{Leonhardt:2006ai}.

\subsection{Spatial Dependent Time Transformation} \label{Sec:SpaceDepTimeTrans}
Next consider a transformation that has also been analysed in more detail from the perspective of creating frequency-altering linear materials, using other methods \cite{Cummer:2010b}, namely
\begin{equation} \label{Eq:SpaceDepTimeTrans}
 T(t',x',y',z')=(t,x,y,z)=\left((ax'+b)t',x',y',z'\right)
\end{equation}
In this case the inverse map is
\begin{equation}
 \mathcal{T}(t,x,y,z)=(t',x',y',z')= \left(\frac{t}{ax+b},x,y,z\right),
\end{equation}
and the Jacobian matrix of $\mathcal{T}$ is
\begin{equation}
 \Lambda\indices{^{\mu}_{\nu}} =
 \begin{pmatrix}
  \frac{1}{ax+b} & -\frac{at}{(ax+b)^2} & 0 & 0\\
  0 & 1 & 0 & 0 \\
  0 & 0 & 1 & 0 \\
  0 & 0 & 0 & 1
 \end{pmatrix}.
\end{equation}
The next step is to calculate $\tilde{\boldsymbol{\chi}}$ using Eqs.\ (\ref{Eq:InitialVacMaterialChi}).  The results can be extracted from identifications with Eq.\ (\ref{Eq:CartesianChi}), and written in the representation of Eq.\ (\ref{Eq:ConstitutiveComponents1}) as
\begin{equation}
\begin{gathered}
 \check{\varepsilon}^* = 
 \begin{pmatrix}
  ax+b & 0 & 0 \\
  0 & \frac{\left((ax+b)^2+at\right)\left((ax+b)^2-at\right)}{(ax+b)^3} & 0 \\
  0 & 0 & \frac{\left((ax+b)^2+at\right)\left((ax+b)^2-at\right)}{(ax+b)^3} 
 \end{pmatrix},
\quad
\check{\mu}^{-1} = \frac{1}{ax+b}
 \begin{pmatrix}
  1 & 0 & 0 \\
  0 & 1 & 0 \\
  0 & 0 & 1
 \end{pmatrix}, \\
\check{\gamma_1}^*=-\check{\gamma_2}^{*\mathtt{T}}= \frac{at}{(ax+b)^2}
 \begin{pmatrix}
  0 & 0 & 0 \\
  0 & 0 & -1 \\
  0 & 1 & 0
 \end{pmatrix}.
\end{gathered}
\end{equation}
When rewritten in the more common representation of Eq.\ (\ref{Eq:ConstitutiveComponents2}) they become
\begin{equation}
 \check{\varepsilon} = \check{\mu} = (ax+b)
 \begin{pmatrix}
  1 & 0 & 0 \\
  0 & 1 & 0 \\
  0 & 0 & 1 
 \end{pmatrix},
\quad
\check{\gamma_1}=\check{\gamma_2}^{\mathtt{T}}= \frac{at}{ax+b}
 \begin{pmatrix}
  0 & 0 & 0 \\
  0 & 0 & -1 \\
  0 & 1 & 0
 \end{pmatrix}.
\end{equation}
This example highlights the slight difference in the meaning of what we are calling $\check{\varepsilon}^*$, $\check{\mu}^{-1}$, $\check{\gamma_1}^*$, and $\check{\gamma_2}^*$ and the interpretation of these quantities based on the representation of Eq.\ (\ref{Eq:ConstitutiveComponents2}).  This difference arises only when non-zero magneto-electric coupling terms are present, and it is easy to switch between the two representations using Eq.\ (\ref{Eq:ConstitutiveShift}).  While the computation and manipulation of large matrices like $\boldsymbol{\chi}$ may see rather daunting, the use of a modern computer algebra package makes the calculations almost trivial.

\subsection{Time Dependent Spatial Transformation} \label{Sec:TimeDepSpaceTrans}
Having looked at the behaviour exhibited by a spatial dependent time transformation, it is natural to next enquire about a time dependent spatial transformation.  Let the transformation be
\begin{equation} \label{Eq:TDepSpaceTrans}
 T(t',x',y',z')=(t,x,y,z)=\left(t',(at'+b)x',y',z'\right)
\end{equation}
The associated map $\mathcal{T}$ is
\begin{equation}
 \mathcal{T}(t,x,y,z)=(t',x',y',z')=\left(t,\frac{x}{at+b},y,z\right),
\end{equation}
and the Jacobian matrix of $\mathcal{T}$ is
\begin{equation}
 \Lambda\indices{^{\mu}_{\nu}} =
 \begin{pmatrix}
  1 & 0 & 0 & 0\\
  -\frac{ax}{(at+b)^2} & \frac{1}{at+b} & 0 & 0 \\
  0 & 0 & 1 & 0 \\
  0 & 0 & 0 & 1
 \end{pmatrix}.
\end{equation}
Proceeding as before and expressing the results in the representation of Eq.\ (\ref{Eq:ConstitutiveComponents2}), 
\begin{equation}
 \check{\varepsilon} = \check{\mu} = \frac{(at+b)^3}{(at+b)^4-a^2x^2}
 \begin{pmatrix}
  \frac{(at+b)^4-a^2x^2}{(at+b)^2} & 0 & 0 \\
  0 & 1 & 0 \\
  0 & 0 & 1 
 \end{pmatrix},
\quad
\check{\gamma_1}=\check{\gamma_2}^{\mathtt{T}}= \frac{ax(at+b)}{(at+b)^4-a^2x^2}
 \begin{pmatrix}
  0 & 0 & 0 \\
  0 & 0 & -1 \\
  0 & 1 & 0
 \end{pmatrix}.
\end{equation}
It is interesting to compare the last few examples.  In the cases of a simple scaling of $t$ versus a spatially dependent transformation of $t$, the resultant materias have similar characteristics in that $\check{\varepsilon}$ is isotropic and scales like $t$, while the mixing of space and time introduces a magneto-electric coupling.  According to the discussion in Sec.\ \ref{Sec:Plebanski}, because the resulting material is isotropic, this magneto-electric coupling, if small enough, could arise from a small material velocity.  So the resulting material could be considered as being either at rest and possessing a magneto-electric coupling, in motion with no magneto-electric coupling, or some combination of the two.

Turning to the cases of a simple scaling of $x$ versus a time dependent transformation of $x$, $\check{\varepsilon}_{xx}$ still scales like $x$, but $\check{\varepsilon}_{yy}=\check{\varepsilon}_{zz}$ have totally different behaviours in each case.  Again the mixing of space and time introduces a magneto-electric coupling, but this time, because the resulting material is not isotropic, these magneto-electric couplings cannot be interpreted as a material velocity.

To verify the material parameters obtained here, a validation calculation must be performed.  Such a calculation should be based on a combined space-time transformation in a complex initial material to demonstrate the full generality of our approach.  Such a calculation, however, requires an in-depth analysis for both the initial medium and the transformed medium that is beyond the scope of this paper, which is intentionally a focused description and formulation of the new idea.  However, such work is in progress.

\subsection{Square Cloak}
To demonstrate agreement with previous results, consider now the well-known, but non-trivial, example of a  square cloak, studied by Rahm et.\ al.\ \cite{Rahm:200887}.  In this example the transformation is a map $T$ piecewise defined by
\begin{equation}
 T(t',x',y',z')=(t,x,y,z)=\left(t,x'\left(\frac{s_2-s_1}{s_2}\right)+s_1,y'\left(\frac{s_2-s_1}{s_2}+\frac{s_1}{x'}\right),z'\right),
\end{equation}
for $(0<x'\leq s_2)$, $(-s_2<y'\leq s_2)$, $|y'|<|x'|$, and $|z'|<\infty$, and $T(t',x',y',z')=(t,x,y,z)$ elsewhere. Again, we are really only interested in the inverse of this map, which we take to be piecewise defined by
\begin{equation}
 \mathcal{T}(t,x,y,z)=(t',x',y',z')=\left(t,\frac{s_2}{s_2-s_1}(x-s_1),\frac{s_2(x-s_1)y}{(s_2-s_1)x},z\right),
\end{equation}
for $(s_1\leq x\leq s_2)$, $(-s_2< y \leq s_2)$ with $|y|<|x|$ and $|z|<\infty$, and $\mathcal{T}(t,x,y,z)=(t',x',y',z')$ for $x>s_2$.  This map is not defined for $x<s_1$, so $\boldsymbol{\chi}$ is undetermined and therefore arbitrary in this region.
The Jacobian matrix of $\mathcal{T}$ is
\begin{equation}
 \Lambda\indices{^{\alpha}_{\beta}} = \begin{pmatrix}
  1 & 0 & 0 & 0 \\
  0 & \frac{s_2}{s_2-s_1} & 0 & 0 \\
  0 & \frac{ys_1s_2}{x^2(s_2-s_1)} & \frac{s_2(x-s_1)}{x(s_2-s_1)} & 0 \\
  0 & 0 & 0 & 1
 \end{pmatrix}
\end{equation}
Turning the crank on Eq.\ (\ref{Eq:InitialVacMaterialChi}), and comparing with Eq.\ (\ref{Eq:CartesianChi}) the components of $\check{\varepsilon}$ and $\check{\mu}$ can be identified and gathered in matrix form as
\begin{equation}
 \check{\varepsilon} = \check{\mu} =
 \begin{pmatrix}
  1-\frac{s_1}{x} & -s_1\frac{y}{x^2} & 0 \\
  -s_1\frac{y}{x^2} & \frac{x^4+s_1^2y^2}{x^3(x-s_1)} & 0\\
  0 & 0 & \frac{s_2^2(x-s_1)}{x(s_1-s_2)^2}
 \end{pmatrix},
\end{equation}
and $\check{\gamma_1}=\check{\gamma_2}=0$, which are exactly the results obtained in Ref.\ \cite{Rahm:200887}.  Note that the domain of $\mathcal{T}$, $(s_1\leq x\leq s_2)$, $(-s_2< y \leq s_2)$ with $|y|<|x|$ and $|z|<\infty$, dictates also the domain of $\check{\varepsilon}=\check{\mu}$, so the material distribution is defined piecewise, with this non-vacuum piece being located exactly where it is desired.

\section{Conclusions} \label{Sec:Conclusions}
Having shown that there may be important limitations to the approach to transformation optics based on the constitutive equations originally presented by Plebanski \cite{Plebanski:1959ff}, we ask whether it is possible to construct a completely covariant and more general approach to transformation optics in linear materials.  The main result of this construction, Eqs.\ (\ref{Eq:MaterialChi}) and (\ref{Eq:InitialVacMaterialChi}), represent the core of transformation optics. 

While the Plebanski based approach remains useful and valid for a wide class of transformations, the benefit of this approach is that it is valid for arbitrary background space-times, arbitrary initial material, arbitrary material motion, and for arbitrary transformations -- including those that mix space and time as.  At first sight the totally covariant approach seems to have the drawback of being more difficult to work with since it involves transforming a matrix with 256 components.  But with the aid of a computer algebra package these kinds of computations become trivial, and the recipe given by Eqs.\ (\ref{Eq:MaterialChi}) or (\ref{Eq:InitialVacMaterialChi}) is straightforward and applicable in all situations. 

Several examples have been discussed that illustrate the material parameter features that might be expected from various types of transformations.  Particularly interesting are the examples of transformations that mix space and time.  It is found that a generic feature of these types of transformations is the appearance of magneto-electric coupling terms.  For transformations that have only a spatially dependent time transformation, the resulting material is isotropic, and the magneto-electric coupling could be reinterpreted as a velocity, in accordance with the Plebanski based method.  However, if the transformation mixes a spatially dependent time function with an additional spatial function of any type, or if the transformation is of a time dependent spatial function type, then the resulting material is not isotropic, and the magneto-electric couplings cannot be reinterpreted as material motion.

\begin{acknowledgments}
The Duke University component of this work was partly supported by a Lockheed Martin University Research Initiative award and by a Multiple University Research Initiative from the Army Research Office (Contract No. W911NF-09-1-0539).
\end{acknowledgments}

\end{document}